%% file: saber.tex
\documentclass[sigconf, screen, nonacm]{cidr-2026}

\usepackage{amsmath,amsfonts}

\usepackage{amssymb}
\usepackage{bm}
\usepackage{booktabs}
\usepackage{enumitem}
\usepackage{multirow}
\usepackage{makecell}
\usepackage{float}
\usepackage{array}
\usepackage{listings}
\usepackage{xcolor}
\usepackage{soul}
\usepackage{tabularx}
\usepackage{adjustbox}
\usepackage{caption}

\makeatletter
\newlength{\ListingContentWidth}
\makeatother

\newcolumntype{L}[1]{>{\raggedright\ttfamily\arraybackslash}p{#1}}

\begin{document}
\title{SABER: A SQL-Compatible Semantic Document Processing System Based on Extended Relational Algebra}

 \author{Changjae Lee}
 \email{changjae@buffalo.edu}
 \affiliation{
   \institution{University at Buffalo}
   \city{}
   \country{}
 }
\author{Zhuoyue Zhao}
 \email{zzhao35@buffalo.edu}
 \affiliation{
   \institution{University at Buffalo}
   \city{}
   \country{}
 }
 \author{Jinjun Xiong}
 \authornote{Corresponding author}
 \email{jinjun@buffalo.edu}
 \affiliation{
   \institution{University at Buffalo}
   \city{}
   \country{}
 }

\input{Body/1_abstract}

\maketitle

\input{Body/2_introduction}
\input{Body/3_related_work}
\input{Body/4_design}
\input{Body/5_implementation}
\input{Body/6_results}
\input{Body/7_conclusion}

\bibliographystyle{ACM-Reference-Format}
\bibliography{references}

\end{document}

%% file: Body/1_abstract.tex
\begin{abstract}
The emergence of large-language models (LLMs) has enabled a new class of semantic data processing systems (SDPSs) to support declarative queries against unstructured documents. Existing SDPSs are, however, lacking a unified algebraic foundation, making their queries difficult to compose, reason, and optimize. We propose a new semantic algebra, SABER (\textbf{S}emantic \textbf{A}lgebra \textbf{B}ased on \textbf{E}xtended \textbf{R}elational algebra), opening the possibility of semantic operations' logical plan construction, optimization, and formal correctness guarantees. We further propose to implement SABER in a SQL-compatible syntax so that it natively supports mixed structured/unstructured data processing. With SABER, we showcase the feasibility of providing a unified interface for existing SDPSs so that it can effectively mix and match any semantically-compatible operator implementation from any SDPS, greatly enhancing SABER's applicability for community contributions. 
\end{abstract}

%% file: Body/2_introduction.tex
\section{Introduction}
\label{sec:introduction}

The emergence of large language models (LLMs) has transformed document-centric data processing~\cite{llmforpaperreview,literaturescreening,lotus,docetl,palimpzest}. LLMs provide rich semantic understanding, enabling systems to interpret, reason over, and extract information from unstructured content using natural language interfaces with low engineering effort.
Building on these capabilities, a new class of \emph{semantic data processing systems} (SDPSs) has emerged. Systems such as LOTUS~\cite{lotus}, DocETL~\cite{docetl}, and Palimpzest~\cite{palimpzest} offer LLM-backed operations for tasks like filtering, extraction, joining, and clustering based on semantic content. These systems expose natural-language-based APIs to support unstructured document processing.

However, we find that a major limitation of the existing SDPSs is the lack of a unified algebraic foundation. Each system defines its own semantic operations in isolation, with nuanced deviations or limitations compared to the standard relational semantics. As a result, multi-step pipelines involving operations such as joins, filtering, grouping, and set/bag difference become difficult to compose, reason about, or optimize. In contrast, the success of SQL systems for structured data is due to the foundation of standard relational algebra~\cite{querylanguagesforbags,databasesystemimplementation,briningordertoqueryoptimization,afoundationforconventionalandtemporalqueryoptimization,fragmentsofbagrelationalalgebra,towardstractablealgebrasforbags,algebraicpropertiesofbagdatatypes,coqmechanised,aformalsemanticsofsqlqueries,amultisetextendedrelationalalgebra}, which enables correct composition and optimization of complex queries.
To bridge this gap, we propose \emph{Semantic Algebra Based on Extended Relational algebra (SABER)}, a formal algebra for semantic data processing that augments standard relational algebra with logical semantic operators with well-defined semantics, such as semantic selection, projection, join, difference, intersection, grouping, and sorting.
Such formalisms are essential not only for correct declarative query evaluation but also for enabling SDPS's logic plan optimization. In this work, we describe how we define the logical semantic operators in SABER. Then we analyze existing SDPSs, including LOTUS, DocETL, and Palimpzest, and demonstrate possible mapping of their operators and their limitations or deviations that prevent them from correctly expressing common queries.
This comparative evaluation reveals critical gaps in functionality, particularly in the support for difference and intersection, which are essential for expressing exclusion and overlap in semantic pipelines. 

Another limitation of existing SDPSs is the inability to handle mixed structured/unstructured data easily due to its laser focus on document processing. In fact, production systems often have to handle a mix or nesting of structured, semi-structured, and unstructured data~\cite{DBLP:journals/corr/OngPV14}.
To address this 
challenge, we augment the SQL syntax with a number of semantic operators (such as  $\texttt{SEM\_SELECT}$, $\texttt{SEM\_WHERE}$, $\texttt{SEM\_ORDER\_BY}$) and implement a query rewriter to compose correct semantic queries in one of the SDPSs using the mapping from the previous analysis. Specifically, the SDPSs' implementations are treated as physical operators that correspond to the logical operators in SABER if they are compatible based on our analysis. Then, our system can freely choose one of the SDPSs' implementations or mix-n-match implementations from different SDPS systems.
For the missing semantically-compatible SDPS operators, SABER falls back to its own implementation.
As a result, we show that SABER opens up an opportunity to provide a unified interface in a system that can integrate existing SDPSs' physical operator implementation, similar to works that provide unified relational interfaces for heterogenous data systems~\cite{10.14778/3236187.3236195,10.1145/3631504.3631510}.

In summary, the major contributions of this work are as follows. (1) We propose SABER as a new SDPS grounded in a semantic algebra based on extended relational algebra, which makes SABER amenable to reasoning its correctness and logic plan optimization.
(2) We propose to implement SABER in a SQL-compatible syntax, making
it directly applicable to processing mixed structured and unstructured data.
(3) Based on our analysis of three modern SDPS systems' semantic compatibility with SABER, we further show that SABER can provide
a unified interface to all SDPSs, allowing SABER to freely reuse semantic implementations from either SDPS systems.
Experimentation further demonstrates the validity and practicality of SABER as a novel SDPS. 

%% file: Body/3_related_work.tex
\section{Related Work}
\label{sec:related-work}

\noindent \textbf{Relational Algebra for SQL.}
Relational algebra serves as the formal backbone of structured query languages, providing a logical framework for reasoning about query correctness, equivalence, and transformation~\cite{querylanguagesforbags,databasesystemimplementation,fragmentsofbagrelationalalgebra,towardstractablealgebrasforbags,algebraicpropertiesofbagdatatypes,coqmechanised}. Classical relational algebra assumes set semantics, but practical database systems predominantly adopt bag (multi-set) semantics to reflect duplicate-preserving behaviors in SQL. To model this, researchers have proposed bag-extended relational algebras with well-defined operators for union, join, difference, and projection. To further capture ordering in query optimization, list-based relational algebra has been introduced~\cite{briningordertoqueryoptimization}, enabling formal treatments of top-$k$ queries.
These algebraic frameworks support algebraic equivalences, transformation rules, and cost-based optimization strategies foundational to SQL query processing.
Lack of such formalism in existing SDPSs presents a key limitation for correctly and efficiently composing and optimizing queries. In this work, we attempt to address the limitation by designing SABER for enabling declarative query evaluation over a mix of structured and unstructured data.

\noindent \textbf{SQL towards Unstructured Data.}
Recent extensions to support unstructured data in SQL include SUQL~\cite{suql}, which introduces \textsc{ANSWER} and \textsc{SUMMARY} as user-defined functions but without algebraic semantics. SSQL~\cite{ssql} enables semantic vector filtering via \texttt{SEMANTIC} clauses, limited by embedding granularity and absence of structured query semantics. UQE~\cite{uqe} defines UQL for unstructured data analytics via LLM-based sampling and planner scheduling. BINDER~\cite{binder} augments symbolic programs with LLM calls through unified APIs, improving generality but lacking algebraic formalism. As they only extend the SQL solely for the specific tasks they are designed for, they cannot be directly used as the common semantic algebra for SDPSs.
In contrast, our work augments the extended relational algebra with a number of logical semantic operators, and our system supports the correct mapping of them to the underlying SDPSs' implementation to preserve SQL semantics.

\noindent \textbf{Semantic Data Processing Systems.}
A growing body of work has investigated LLM-backed semantic data processing. Systems such as LOTUS~\cite{lotus}, DocETL~\cite{docetl}, and Palimpzest~\cite{palimpzest} exemplify this trend, each introducing abstractions for semantic filtering, extraction, joining, or clustering based on LLM-generated embeddings or prompt responses.
While these systems demonstrate the practical benefits of integrating LLMs into data workflows, they lack a shared formal semantic foundation. Operator semantics are defined in system-specific terms, without unified algebraic rules or formal notions of equivalence. Additionally, the supported sets of operators vary across systems, and core semantic relational primitives such as difference, intersection, or deduplication are often missing or only partially implemented.
To date, no existing SDPS defines a general-purpose algebra that integrates LLM-driven semantics with SQL-compatible operators. This gap limits composability, interoperability, and formal reasoning, motivating the need for a framework like SABER to unify semantic and relational paradigms.

\noindent \textbf{Cross-platform data processing.}
Apache Wayang \cite{10.14778/3236187.3236195,10.1145/3631504.3631510} is a unified framework for integrating multiple data systems. It allows mix-n-match physical operator implementation from different data systems to compose a query pipeline for a SQL query under the standard extended relational algebra. Different from Apache Wayang, our work extends the relational algebra with LLM-backed semantic operators and enables the integration of SDPSs.

%% file: Body/4_design.tex
\section{The Design of SABER}
\label{sec:design}

\subsection{Algebraic Form and Integration Potential}

We present \emph{Semantic Algebra Based on Extended Relational Algebra
(SABER)}, which extends the extended relational algebra framework
introduced by~\cite{briningordertoqueryoptimization}, where relations
are represented as {ordered} lists to capture duplicates and ordering, moving beyond the set-theoretic formulation. Additional insights are drawn from~\cite{databasesystemimplementation,querylanguagesforbags,amultisetextendedrelationalalgebra}.

Table~\ref{tab:full-ra-sql-saber} provides an overview of SABER. It categorizes operators into three groups: \textit{Basic}, which includes fundamental relational expressions~\cite{amultisetextendedrelationalalgebra}; \textit{Compound}, which consists of operators expressible using the basic ones; and \textit{Extended}, which contains operators that support some of the additional features of SQL (e.g., \texttt{ORDER BY}) that cannot be expressed using only Basic and Compound operators. 
SABER comprises 12 conventional relational algebra operators and 10 semantic relational operators. Semantic operators are denoted with a superscript \textit{sem}, e.g., $\sigma^{\textit{sem}}$, to distinguish them from their classical counterparts. Each semantic operator in SABER incorporates language-based reasoning, similarity computation, or prompt-driven transformations to handle unstructured or loosely structured data.
The Product, {Bag-Union, which corresponds to Union-all in \cite{briningordertoqueryoptimization}}, and Top-$k$ operators do not have semantic counterparts because their $\lambda$-calculus definitions rely solely on standard auxiliary functions and the $Loop$ function~\cite{briningordertoqueryoptimization}. 
In other words, unlike other relational algebra operators, they do not need LLM-based semantic involvement.

Formally, the semantic operators are defined by replacing the non-semantic data transformation or comparison with semantic transformation or comparison in their counter-parts in the standard extended relational algebra. For example, $\sigma_P^{sem} (r)$ is defined as filtering each row from subexpression $r$ by evaluating the boolean semantic predicate $P$ on it, which is usually implemented by invoking LLM with a prompt comprising the user provided natural language predicate augmented with additional metadata and prompt words, and only retain those where $P$ evaluates to true. Deduplication $\delta^{sem}$ is defined based on similarity-based equality instead of data type-based equality.
Semantic projection is defined over a collection $\mathcal{F}^{\textit{sem}}$ of semantic expressions ${{{f}}_i}^{\textit{sem}}$.
Semantic aggregation is defined over a function set $\mathbb{F}^{\textit{sem}}$ containing operators $F_i^{\textit{sem}}$ with semantic interpretation. Semantic grouping---corresponding to \texttt{GetGroup} in~\cite{briningordertoqueryoptimization}---generalizes equality-based grouping by collecting all tuples whose group-by attributes are semantically equivalent to those of the reference tuple. Semantic difference
relies on ${\texttt{isIn}}^{\textit{sem}}$, which evaluates membership based on semantic equivalence. Semantic sorting uses an attribute order specification $a^{\textit{sem}}$ that reflects semantic comparability rather than syntactic order. 

\begin{table*}[h]
\centering
\small
\renewcommand{\arraystretch}{1.15}
\caption{Overview of SABER}
\label{tab:full-ra-sql-saber}
\begin{tabular}{ll>{\centering\arraybackslash}p{1cm}p{2cm}>{\centering\arraybackslash}p{1cm}p{8.4cm}}
\toprule
\textbf{Category} & \textbf{Operator Name} & \shortstack[c]{\textbf{RA} \\ \textbf{Symbol}} & \textbf{SQL Mapping} & \shortstack[c]{\textbf{Semantic} \\ \textbf{Operator}} & \textbf{SABER SQL UDF} \\
\midrule
\multirow{7}{*}{Basic}
  & Selection           & $\sigma$       & \texttt{WHERE}         & $\sigma^{\textit{sem}}$        & \texttt{SEM\_WHERE(`semantic\_query')} \\
  & Projection          & $\pi$          & \texttt{SELECT}        & $\pi^{\textit{sem}}$           & \texttt{SEM\_SELECT(`semantic\_query') AS alias} \\
  & Product             & $\times$       & Relations in \texttt{FROM}          & \texttt{N/A}                            & \texttt{N/A} \\
  & Set-Difference      & $-_{S}$        & \texttt{EXCEPT}        & $-_{S}^{\textit{sem}}$         & \texttt{SEM\_DISTINCT(SEM\_EXCEPT\_ALL(SABER query1, SABER query2))} \\
  & Bag-Difference      & $-_{B}$        & \texttt{EXCEPT ALL}    & $-_{B}^{\textit{sem}}$         & \texttt{SEM\_EXCEPT\_ALL(SABER query1, SABER query2)} \\
	& Set-Union           & $\cup_{S}$     & \texttt{UNION}         & $\cup_{S}^{\textit{sem}}$       & \texttt{SEM\_DISTINCT(SABER query1 UNION ALL SABER query2)} \\
  & Bag-Union           & $\cup_{B}$     & \texttt{UNION ALL}     & \texttt{N/A}                            & \texttt{N/A} \\
\midrule
\multirow{3}{*}{Compound}
  & Set-Intersection    & $\cap_{S}$     & \texttt{INTERSECT}     & $\cap_{S}^{\textit{sem}}$      & \texttt{SEM\_DISTINCT(SEM\_INTERSECT\_ALL(SABER query1, SABER query2))} \\
  & Bag-Intersection    & $\cap_{B}$     & \texttt{INTERSECT ALL} & $\cap_{B}^{\textit{sem}}$      & \texttt{SEM\_INTERSECT\_ALL(SABER query1, SABER query2)} \\
  & Join                & $\Join$        & \texttt{JOIN}          & $\Join^{\textit{sem}}$         & \texttt{SEM\_JOIN(Table1, Table2, `semantic\_query')}  \\
\midrule
\multirow{5}{*}{Extended}
  & Grouping            & $\gamma$       & \texttt{GROUP BY}      & $\gamma^{\textit{sem}}$        & \texttt{SEM\_GROUP\_BY(attribute, k)} \\
  & Aggregation         & $\xi$          & \texttt{SUM}, \texttt{AVG}, etc. & $\xi^{\textit{sem}}$ & \texttt{SEM\_AGG([attribute, ]`semantic\_query') AS alias} \\
  & Deduplication       & $\delta$       & \texttt{DISTINCT}      & $\delta^{\textit{sem}}$        & \texttt{SEM\_DISTINCT(attribute)} \\
  & Sorting             & $\tau$         & \texttt{ORDER BY}      & $\tau^{\textit{sem}}$          & \texttt{SEM\_ORDER\_BY([attribute, ]`semantic\_query')} \\
  & Top-$k$             & $\lambda$      & \texttt{LIMIT}         & \texttt{N/A}                            & \texttt{N/A} \\
\bottomrule
\end{tabular}
\end{table*}

Semantic operators have the same semantics as their relational algebra counterparts, with the only difference being that the data transformation/predicate (which are type-checked black boxes to relational algebra) are implemented in LLM rather than traditional programming languages. As a result, transformation rules defined for conventional relational algebra (e.g., selection push-down, projection composition, duplicate elimination propagation) are applicable to semantic operators as well. That is, if a rule $e_1 \equiv e_2$ holds in the conventional setting, the corresponding semantic rule $e_1^{\textit{sem}} \equiv^{\textit{sem}} e_2^{\textit{sem}}$ can be applied under the appropriate semantic equivalence.

By grounding semantic processing in a formal algebra compatible with SQL's extended semantics, SABER serves as a bridge between the structured and unstructured data processing paradigms. This enables systematic integration of LLM-powered semantic transformations into the relational model and paves the way for principled optimization, hybrid reasoning, and formal semantics in next-generation data systems.

\subsection{Comparison of SABER and Existing SDPS}

Grounded in the SABER framework introduced above, we systematically analyze three representative SDPSs---LOTUS~\cite{lotus}, DocETL~\cite{docetl}, and Palimpzest~\cite{palimpzest}---to assess their operational coverage. These systems exemplify cutting-edge approaches to LLM-integrated document analysis, each offering different abstractions and pipelines for semantic data manipulation.

We evaluate each system in terms of its support for SABER semantic operators (Table~\ref{tab:coverage}). 
We categorize the operators into \textbf{Basic}, \textbf{Compound}, and \textbf{Extended} as Table~\ref{tab:full-ra-sql-saber}, and annotate support via documented APIs and system behavior.
Table~\ref{tab:coverage} reveals both commonalities and divergences among the three systems:

\begin{itemize}
    \item All three systems support semantic \textbf{selection} ($\sigma^{\textit{sem}}$) and \textbf{projection} ($\pi^{\textit{sem}}$), reflecting their core role in LLM-driven filtering and transformation.
    \item Semantic \textbf{join} ($\Join^{\textit{sem}}$){\footnote{We
		only consider equi-join for $\Join^{\textit{sem}}$ for now
		and leave general $\theta$ join for future work.}} is supported in LOTUS and DocETL, but not in Palimpzest, highlighting divergence in pipeline composability.
    \item Extended operators such as \textbf{grouping} ($\gamma^{\textit{sem}}$), \textbf{aggregation} ($\xi^{\textit{sem}}$), \textbf{deduplication} ($\delta^{\textit{sem}}$), and \textbf{sorting} ($\tau^{\textit{sem}}$) are variably supported. Palimpzest offers only partial aggregation, whereas LOTUS and DocETL provide richer operator sets.
    \item Critically, none of the systems supports semantic
		\textbf{difference} ($-^{\textit{sem}}$) or
		\textbf{intersection} ($\cap^{\textit{sem}}$) directly, making
		it challenging to construct these
		operators.  For instance, semantic difference cannot be
		expressed in these systems. For $\cap^{\textit{sem}}$, it is
		not natively supported by any of the three SDPSs -- it can
		only be composed through a combination of $\pi^{\textit{sem}}$ and $\Join^{\textit{sem}}$: all attributes are first projected into a single attribute using $\pi^{\textit{sem}}$, after which applying $\Join^{\textit{sem}}$ yields a result that is semantically equivalent to $\cap^{\textit{sem}}$. These omissions are significant, as both operations are essential for capturing exclusion and overlapping patterns in comparative and conditional analyses.
\end{itemize}

\begin{table*}[h]
  \centering
  \small
  \renewcommand{\arraystretch}{1.15}
  \setlength{\tabcolsep}{4pt}
  \begin{tabular}{@{\hskip 2pt}ll@{\hskip 6pt}c@{\hskip 6pt}c@{\hskip 6pt}c@{\hskip 2pt}}
        \toprule
    \textbf{Category} & \textbf{Operator} & \textbf{LOTUS} & \textbf{DocETL} & \textbf{Palimpzest} \\
    \midrule
    \multirow{3}{*}{\textbf{Basic}} 
      & $\sigma^{\textit{sem}}$ Selection & 
        \checkmark\,\makecell[l]{\texttt{sem\_filter}} & 
        \checkmark\,\makecell[l]{\texttt{Filter}} & 
        \checkmark\,\makecell[l]{\texttt{sem\_filter},\\\texttt{filter}} \\
      & $\pi^{\textit{sem}}$ Projection & 
        \checkmark\,\makecell[l]{\texttt{sem\_map},\\\texttt{sem\_extract}} & 
        \checkmark\,\makecell[l]{\texttt{Map}, \texttt{Parallel Map},\\\texttt{Extract}} & 
        \checkmark\,\makecell[l]{\texttt{sem\_add\_columns},\\\texttt{project}, \texttt{map}} \\
      & $-^{\textit{sem}}$ Difference & 
        $\times$ & $\times$ & $\times$ \\
    \midrule
    \multirow{2}{*}{\textbf{Compound}} 
      & $\cap^{\textit{sem}}$ Intersection & 
        $\times$ & $\times$ & $\times$ \\
      & $\Join^{\textit{sem}}$ Join & 
        \checkmark\,\makecell[l]{\texttt{sem\_join},\\\texttt{sem\_sim\_join}} & 
        \checkmark\,\makecell[l]{\texttt{Equijoin}} & 
        $\times$ \\
    \midrule
    \multirow{4}{*}{\textbf{Extended}} 
      & $\gamma^{\textit{sem}}$ Group-by & 
        \checkmark\,\makecell[l]{\texttt{sem\_cluster\_by}} & 
        \checkmark\,\makecell[l]{\texttt{Cluster}} & 
        \checkmark\,\makecell[l]{\texttt{groupby}} \\
      & $\xi^{\textit{sem}}$ Aggregation & 
        \checkmark\,\makecell[l]{\texttt{sem\_agg}} & 
        \checkmark\,\makecell[l]{\texttt{Reduce}} & 
        $\triangle$\,\makecell[l]{\texttt{count}, \texttt{average}} \\
      & $\delta^{\textit{sem}}$ Deduplication & 
        \checkmark\,\makecell[l]{\texttt{sem\_dedup}} & 
        \checkmark\,\makecell[l]{\texttt{Resolve}} & 
        $\times$ \\
      & $\tau^{\textit{sem}}$ Sorting & 
        \checkmark\,\makecell[l]{\texttt{sem\_topk}} & 
        \checkmark\,\makecell[l]{\texttt{Rank}} & 
        \checkmark\,\makecell[l]{\texttt{retrieve}} \\
    \bottomrule
  \end{tabular}
  \caption{Support for Semantically Enriched Relational Algebra Operators Across SDPSs}
  \label{tab:coverage}
\end{table*}

SABER thus serves not only as a blueprint for identifying such gaps but also as a guide for actionable system evolution. By formalizing semantic operator semantics within an algebraic framework, SDPSs can achieve the same advantages that SQL systems have long leveraged: composability, rewrite rules, and query plan optimizations grounded in well-defined operator semantics.

%% file: Body/5_implementation.tex
\section{SQL-Based Implementation of SABER}
\label{sec:implementation}

\subsection{System Architecture and Workflow}

We implement the Semantic Algebra Based on Extended Relational algebra (SABER) through SQL-accessible UDF-style interfaces that integrate semantic reasoning into structured queries. Our architecture separates relational execution from semantic evaluation, yet keeps both interoperable under a unified SQL front-end.

Internally, semantic operations are triggered not through SQL parsing or logical plan transformations---as is standard in classical relational engines---but through pattern matching against SQL strings using regular expressions. This pragmatic approach avoids invasive changes to the SQL parser and enables rapid deployment in existing systems. Each semantic operator invocation (e.g., \texttt{SEM\_JOIN(...)}) is identified, parsed, and dispatched to dedicated runtime handlers that implement the corresponding SABER semantics.

The system follows a three-stage pipeline:
\begin{enumerate}
  \item \textbf{SQL String Pattern Matching}: The input SQL is scanned using regex patterns to identify and extract semantic UDF invocations and their arguments.
  \item \textbf{Semantic Execution}: For each matched operator, the system loads the referenced data (e.g., from tables or subqueries), applies the appropriate LLM-backed transformation (e.g., prompt-based projection), and materializes the output as an intermediate table or dataframe.
  \item \textbf{Hybrid Reassembly}: The modified SQL query is rewritten to substitute semantic calls with references to the materialized outputs, enabling standard relational engines to continue processing.
\end{enumerate}

The following SABER query exemplifies this pipeline in action. It answers the natural language question: ``Among products, what is the most expensive apple-related one?'' Here, the \texttt{SEM\_WHERE} clause triggers semantic execution, where product names are semantically filtered based on their relation to ``apple'':

\begin{lstlisting}[language=SQL]
SELECT name, price 
FROM products 
WHERE SEM_WHERE('{name} is related to apple', 'lotus') 
ORDER BY price DESC 
LIMIT 1;
\end{lstlisting}

We implement SABER on top of three existing SDPSs rather than building a new system from scratch. As these SDPSs do not support semantic difference and intersection, we implement them using embedding-based similarity. This approach allows us to benefit from existing query optimization while showcasing the flexibility of our design. The architecture remains modular and non-intrusive: it isolates the semantic runtime and supports hybrid pipeline execution without requiring changes to SQL parsers or query optimizers, leaving deeper integration and optimization of custom semantics as directions for future work.

\subsection{Semantic SQL UDF Syntax}

Table~\ref{tab:full-ra-sql-saber} shows our UDF-style syntax for invoking SABER operators. Each semantic operator is denoted with a superscript $\textit{sem}$ and implemented as a standalone Python function internally matched and executed. Optional arguments such as prompts or system templates allow flexible interaction with LLM-based semantics.

\subsection{Composability and Query Expressiveness}

The regex-based operator extraction mechanism allows SABER UDFs to interleave seamlessly with classical SQL clauses, enabling hybrid queries that operate over both relational tables and LLM-interpreted document structures. The system supports modularity by encapsulating each semantic UDF as a self-contained logical transformation grounded in SABER semantics. It also offers flexibility through prompt and template parameters that dynamically steer semantic behavior, encouraging re-usability and experimentation. Despite leveraging LLMs, the UDFs preserve SQL's declarative nature by abstracting away model-specific operations.

Crucially, the design maintains algebraic closure: the outputs of semantic UDFs are relational tables that remain compatible with downstream SQL operators. This paves the way for future extensions, including cost-based planning that spans semantic and classical operators, intermediate materialization strategies using vector caches or partial execution, and formal provenance tracking of semantically transformed data.

Overall, this implementation bridges relational and semantic paradigms through a pragmatic SQL-first interface, laying the foundation for fully integrated hybrid query engines.

%% file: Body/6_results.tex
\section{Experimental Results}
\label{sec:results}

We evaluate SABER's expressiveness and utility by executing a semantically enriched SQL query over a real dataset using three representative SDPS backends: LOTUS, DocETL, and Palimpzest. Although these systems interface with LLMs through distinct modalities, none supports algebraically composable semantics. SABER addresses this limitation by embedding declarative, operator‐based semantic constructs directly into SQL.

\subsection{Task and Data}

This experiment is driven by the following natural‐language query:

\begin{quote}
\textit{What are the top 5 rated movies about personal resilience that were directed by directors who overcame significant personal challenges?}
\end{quote}

We construct a normalized semantic database by integrating IMDb metadata from the official IMDb non-commercial dataset\footnote{\url{https://developer.imdb.com/non-commercial-datasets/}} and the Cinemagoer (IMDbPY) library\footnote{\url{https://github.com/cinemagoer/cinemagoer}}. The integration process begins by extracting mappings between movies and their directors from the \texttt{title.crew.tsv.gz} file. We then retrieve the top 250 movies via the IMDbPY API, collecting metadata such as title, year, rating, and plot. Each movie is linked to its primary director to ensure relational uniqueness. Biographical information for each director---including summaries and personal histories---is also fetched via IMDbPY. The resulting schema consists of two relational tables: \texttt{movies}, which contains film-level metadata (e.g., \texttt{tconst}, title, rating, plot), and \texttt{directors}, which includes director-specific contextual information (e.g., \texttt{nmconst}, \texttt{name}, \texttt{biography}). For instance, the entry for \textit{The Shawshank Redemption} (rated 9.3) is linked to director Frank Darabont, whose biography includes his experiences as a refugee and early writing struggles.

\begin{figure}[t]
  \centering
  % LOTUS
  \begin{minipage}[t]{\linewidth}
    % \textbf{LOTUS}
    \begin{lstlisting}[language=SQL]
SELECT m.title, d.name AS director, m.year, m.rating,
       SEM_SELECT('Summarize biography of the director related to overcoming challenges in one short sentence.') AS director_summary,
FROM movies AS m JOIN directors AS d ON m.nmconst = d.nmconst
WHERE SEM_WHERE('the director overcame significant personal challenges.') AND 
      SEM_WHERE('the plot is about personal resilience.')
ORDER BY CAST(m.rating AS FLOAT) DESC
LIMIT 5;
    \end{lstlisting}
  \end{minipage}
  \captionof{figure}{Backend-free SABER SQL query}
	\vspace{-5mm}
  \label{fig:unified_queries}
\end{figure}

\begin{figure*}[t]
  \centering
  % LOTUS
  \begin{minipage}[t]{0.30\textwidth}
    \textbf{LOTUS}
    \begin{lstlisting}[language=SQL]
SELECT m.title, d.name AS director, m.year, m.rating,
       SEM_SELECT('Summarize {d.biography} focusing on overcoming challenges in a single sentence', 'lotus') AS director_summary,
FROM movies AS m JOIN directors AS d ON m.nmconst = d.nmconst
WHERE SEM_WHERE('{d.biography} highlights overcoming significant personal challenges', 'lotus') AND 
      SEM_WHERE('{m.plot} describes personal resilience', 'lotus')
ORDER BY CAST(m.rating AS FLOAT) DESC
LIMIT 5;
    \end{lstlisting}
  \end{minipage}\hfill
  % DocETL
  \begin{minipage}[t]{0.36\textwidth}
    \textbf{DocETL}
    \begin{lstlisting}[language=SQL]
SELECT m.title, d.name AS director, m.year, m.rating,
       SEM_SELECT('Director Biography: {{ input.d.biography }}

Summarize the directors biography focusing on how they overcame challenges in one short sentence.', 'docetl') AS director_summary,
FROM movies AS m JOIN directors AS d ON m.nmconst = d.nmconst
WHERE SEM_WHERE('Director Biography: {{ input.d.biography }}

Analyze this biography to determine if the director overcame significant personal challenges and return True or False.', 'docetl') AND 
      SEM_WHERE('Movie Plot: {{ input.m.plot }}

Analyze if the plot is about personal resilience and return True or False.', 'docetl')
ORDER BY CAST(m.rating AS FLOAT) DESC
LIMIT 5;
    \end{lstlisting}
  \end{minipage}\hfill
  % Palimpzest
  \begin{minipage}[t]{0.30\textwidth}
    \textbf{Palimpzest}
    \begin{lstlisting}[language=SQL]
SELECT m.title, d.name AS director, m.year, m.rating,
       SEM_SELECT('Summarize biography of the director related to overcoming challenges in one short sentence.', 'palimpzest') AS director_summary,
FROM movies AS m JOIN directors AS d ON m.nmconst = d.nmconst
WHERE SEM_WHERE('the director overcame significant personal challenges', 'palimpzest') AND
      SEM_WHERE('the plot is about personal resilience', 'palimpzest')
ORDER BY CAST(m.rating AS FLOAT) DESC
LIMIT 5;
    \end{lstlisting}
  \end{minipage}
  \caption{Backend‐specific SABER SQL queries}
	% \vspace{-5mm}
  \label{fig:queries}
\end{figure*}

\begin{table*}[h]
\centering
\scriptsize
\renewcommand{\arraystretch}{1.05}
\begin{tabular*}{\textwidth}{@{\extracolsep{\fill}} l|l l r r p{8cm} }
\hline
Backend & Title & Director & Year & Rating & Director Summary \\
\hline
\multirow{5}{*}{LOTUS}
 & The Shawshank Redemption & Frank Darabont & 1994 & 9.3 & Frank Darabont, born in a refugee camp in France and raised in Los Angeles, overcame... \\
 & One Flew Over the Cuckoo's Nest & Milos Forman & 1975 & 8.7 & Milos Forman, orphaned during World War II after losing his parents to the Nazis, overcame... \\
 & The Pianist & Roman Polanski & 2002 & 8.5 & Roman Polanski, a Polish filmmaker born in 1933, overcame the harrowing challenges... \\
 & Modern Times & Charles Chaplin & 1936 & 8.5 & Charlie Chaplin overcame numerous challenges throughout his life, including a tumultuous... \\
 & Bicycle Thieves & Vittorio De Sica & 1948 & 8.2 & Vittorio De Sica overcame the challenges of a poor upbringing in Naples by transitioning... \\
\hline
\multirow{5}{*}{DocETL}
 & The Shawshank Redemption & Frank Darabont & 1994 & 9.3 & Frank Darabont overcame the challenges of being a refugee child by establishing himself... \\
 & One Flew Over the Cuckoo's Nest & Milos Forman & 1975 & 8.7 & Milos Forman surmounted the traumatic loss of his parents during World War II and the... \\
 & City of God & Fernando Meirelles & 2002 & 8.6 & Fernando Meirelles overcame the challenges of transforming a complex story with over... \\
 & The Pianist & Roman Polanski & 2002 & 8.5 & Roman Polanski overcame immense challenges during his childhood, including surviving the... \\
 & American History X & Tony Kaye & 1998 & 8.5 & Tony Kaye faced significant challenges in his career, including disowning the final cut of... \\
\hline
\multirow{5}{*}{Palimpzest}
 & The Shawshank Redemption & Frank Darabont & 1994 & 9.3 & Frank Darabont overcame the challenges of being a refugee and struggling in the film... \\
 & One Flew Over the Cuckoo's Nest & Milos Forman & 1975 & 8.7 & Milos Forman overcame the loss of his parents during World War II and political upheaval... \\
 & It's a Wonderful Life & Frank Capra & 1946 & 8.6 & Frank Capra overcame poverty, family opposition to his education, and professional... \\
 & Harakiri & Masaki Kobayashi & 1962 & 8.6 & Masaki Kobayashi overcame the challenge of being a prisoner of war to create impactful... \\
 & The Lion King & Roger Allers & 1994 & 8.5 & Roger Allers overcame the challenge of having his project 'Kingdom of the Sun' retooled... \\
\hline
\end{tabular*}
\caption{Top Movies per Backend Related to Personal Resilience (Director Summaries Truncated for Space)}
\label{tab:movie-results}
\end{table*}

We issue the SABER query shown in Figure~\ref{fig:unified_queries}.
The SABER query combines semantic projection ($\pi^{\textit{sem}}$)
and selection ($\sigma^{\textit{sem}}$) operators in a unified
execution plan:
\begin{itemize}
  \item $\pi^{\textit{sem}}$ (\texttt{SEM\_SELECT}) summarizes directors' biography in one sentence.
  \item $\sigma^{\textit{sem}}$ (\texttt{SEM\_WHERE}) filters to include only directors who overcame significant personal challenges.
\end{itemize}

\subsection{SABER Query Rewriting}

Our system can rewrite the unified query for each of the three backends, which instantiates these operators with its own LLM prompt. Figure~\ref{fig:queries} shows the rewritten SQL queries.

\subsection{System Comparison and Discussion}

Table~\ref{tab:movie-results} shows the results from each backend. While each system employs its own LLM prompting schema, SABER abstracts these differences and ensures consistent semantic interpretation.

The results demonstrate SABER's capacity to encapsulate LLM‐driven transformations within declarative algebra, enabling structured reasoning over descriptive fields and ensuring portable semantic intent across diverse backends. This unified SQL paradigm supports systematic comparison of semantic query implementations and lays the groundwork for extensible operator‐based semantics in SDPS architectures.

%% file: Body/7_conclusion.tex
\section{Conclusion and Future Work}
\label{sec:conclusion}

We have introduced \emph{Semantic Algebra Based on Extended Relational algebra (SABER)}, a principled algebraic framework that integrates semantic processing into SQL-based data systems. SABER extends classical relational algebra with LLM-backed semantic counterparts, supporting semantic selection, projection, difference, intersection, join, group-by, aggregation, deduplication, and sorting. By embedding these operators as user-defined SQL functions, SABER preserves SQL's declarative nature while substantially enhancing its expressiveness over unstructured and semi-structured data.

SABER's algebraic operators unify and expose existing semantic functionalities from multiple systems---LOTUS, DocETL, and Palimpzest---under a common relational abstraction. Our SQL-based interface enables these operators to be invoked as UDFs, supporting hybrid pipelines that seamlessly integrate structured querying with LLM-driven semantics. Experiments across these systems demonstrate SABER's portability, compositionality, and expressive power.

Looking ahead, several promising directions arise:
\begin{itemize}
  \item \textbf{Formal Semantics for LLM Operators.} While SABER defines operator-level behavior algebraically, the underlying LLM responses—such as prompt interpretation and similarity scoring—lack formal guarantees. Developing probabilistic or approximate models of these behaviors is an important direction for future work.
  \item \textbf{Native Integration with Query Engines.} The current prototype relies on regex-based UDF dispatch. Extending SQL parsers and query planners to natively support SABER semantics would enable deeper optimization and execution efficiency.
  \item \textbf{Semantic Query Optimization.} SABER presently lacks cost-based reasoning and semantic-aware plan rewriting. Future work will explore integrating semantic operators into logical and physical query optimizers to support scalable execution.
\end{itemize}

We view SABER as a foundational step toward a unified, declarative framework for semantic data processing. Future efforts will investigate hybrid optimization strategies, formal verification of semantic operator behavior, and principled extensions to additional modalities such as vision and multi-modal tables.

%% file: saber.bbl
%%% -*-BibTeX-*-
%%% Do NOT edit. File created by BibTeX with style
%%% ACM-Reference-Format-Journals [18-Jan-2012].

\begin{thebibliography}{22}

%%% ====================================================================
%%% NOTE TO THE USER: you can override these defaults by providing
%%% customized versions of any of these macros before the \bibliography
%%% command.  Each of them MUST provide its own final punctuation,
%%% except for \shownote{} and \showURL{}.  The latter two
%%% do not use final punctuation, in order to avoid confusing it with
%%% the Web address.
%%%
%%% To suppress output of a particular field, define its macro to expand
%%% to an empty string, or better, \unskip, like this:
%%%
%%% \newcommand{\showURL}[1]{\unskip}   % LaTeX syntax
%%%
%%% \def \showURL #1{\unskip}           % plain TeX syntax
%%%
%%% ====================================================================

\ifx \showCODEN    \undefined \def \showCODEN     #1{\unskip}     \fi
\ifx \showISBNx    \undefined \def \showISBNx     #1{\unskip}     \fi
\ifx \showISBNxiii \undefined \def \showISBNxiii  #1{\unskip}     \fi
\ifx \showISSN     \undefined \def \showISSN      #1{\unskip}     \fi
\ifx \showLCCN     \undefined \def \showLCCN      #1{\unskip}     \fi
\ifx \shownote     \undefined \def \shownote      #1{#1}          \fi
\ifx \showarticletitle \undefined \def \showarticletitle #1{#1}   \fi
\ifx \showURL      \undefined \def \showURL       {\relax}        \fi
% The following commands are used for tagged output and should be
% invisible to TeX
\providecommand\bibfield[2]{#2}
\providecommand\bibinfo[2]{#2}
\providecommand\natexlab[1]{#1}
\providecommand\showeprint[2][]{arXiv:#2}

\bibitem[Agrawal et~al\mbox{.}(2018)]%
        {10.14778/3236187.3236195}
\bibfield{author}{\bibinfo{person}{Divy Agrawal}, \bibinfo{person}{Sanjay Chawla}, \bibinfo{person}{Bertty Contreras-Rojas}, \bibinfo{person}{Ahmed Elmagarmid}, \bibinfo{person}{Yasser Idris}, \bibinfo{person}{Zoi Kaoudi}, \bibinfo{person}{Sebastian Kruse}, \bibinfo{person}{Ji Lucas}, \bibinfo{person}{Essam Mansour}, \bibinfo{person}{Mourad Ouzzani}, \bibinfo{person}{Paolo Papotti}, \bibinfo{person}{Jorge-Arnulfo Quian\'{e}-Ruiz}, \bibinfo{person}{Nan Tang}, \bibinfo{person}{Saravanan Thirumuruganathan}, {and} \bibinfo{person}{Anis Troudi}.} \bibinfo{year}{2018}\natexlab{}.
\newblock \showarticletitle{RHEEM: enabling cross-platform data processing: may the big data be with you!}
\newblock \bibinfo{journal}{\emph{Proc. VLDB Endow.}} \bibinfo{volume}{11}, \bibinfo{number}{11} (\bibinfo{date}{July} \bibinfo{year}{2018}), \bibinfo{pages}{1414–1427}.
\newblock
\showISSN{2150-8097}
\href{https://doi.org/10.14778/3236187.3236195}{doi:\nolinkurl{10.14778/3236187.3236195}}


\bibitem[Albert(1991)]%
        {algebraicpropertiesofbagdatatypes}
\bibfield{author}{\bibinfo{person}{Joseph Albert}.} \bibinfo{year}{1991}\natexlab{}.
\newblock \showarticletitle{Algebraic Properties of Bag Data Types}. In \bibinfo{booktitle}{\emph{Proceedings of the 17th International Conference on Very Large Data Bases}} \emph{(\bibinfo{series}{VLDB '91})}. \bibinfo{publisher}{Morgan Kaufmann Publishers Inc.}, \bibinfo{address}{San Francisco, CA, USA}, \bibinfo{pages}{211–219}.
\newblock
\showISBNx{1558601503}


\bibitem[Beedkar et~al\mbox{.}(2023)]%
        {10.1145/3631504.3631510}
\bibfield{author}{\bibinfo{person}{Kaustubh Beedkar}, \bibinfo{person}{Bertty Contreras-Rojas}, \bibinfo{person}{Haralampos Gavriilidis}, \bibinfo{person}{Zoi Kaoudi}, \bibinfo{person}{Volker Markl}, \bibinfo{person}{Rodrigo Pardo-Meza}, {and} \bibinfo{person}{Jorge-Arnulfo Quian\'{e}-Ruiz}.} \bibinfo{year}{2023}\natexlab{}.
\newblock \showarticletitle{Apache Wayang: A Unified Data Analytics Framework}.
\newblock \bibinfo{journal}{\emph{SIGMOD Rec.}} \bibinfo{volume}{52}, \bibinfo{number}{3} (\bibinfo{date}{Nov.} \bibinfo{year}{2023}), \bibinfo{pages}{30–35}.
\newblock
\showISSN{0163-5808}
\href{https://doi.org/10.1145/3631504.3631510}{doi:\nolinkurl{10.1145/3631504.3631510}}


\bibitem[Benzaken and Contejean(2019)]%
        {coqmechanised}
\bibfield{author}{\bibinfo{person}{V\'{e}ronique Benzaken} {and} \bibinfo{person}{\'{E}velyne Contejean}.} \bibinfo{year}{2019}\natexlab{}.
\newblock \showarticletitle{A Coq mechanised formal semantics for realistic SQL queries: formally reconciling SQL and bag relational algebra}. In \bibinfo{booktitle}{\emph{Proceedings of the 8th ACM SIGPLAN International Conference on Certified Programs and Proofs}} (Cascais, Portugal) \emph{(\bibinfo{series}{CPP 2019})}. \bibinfo{publisher}{Association for Computing Machinery}, \bibinfo{address}{New York, NY, USA}, \bibinfo{pages}{249–261}.
\newblock
\showISBNx{9781450362221}
\href{https://doi.org/10.1145/3293880.3294107}{doi:\nolinkurl{10.1145/3293880.3294107}}


\bibitem[Cheng et~al\mbox{.}(2023)]%
        {binder}
\bibfield{author}{\bibinfo{person}{Zhoujun Cheng}, \bibinfo{person}{Tianbao Xie}, \bibinfo{person}{Peng Shi}, \bibinfo{person}{Chengzu Li}, \bibinfo{person}{Rahul Nadkarni}, \bibinfo{person}{Yushi Hu}, \bibinfo{person}{Caiming Xiong}, \bibinfo{person}{Dragomir Radev}, \bibinfo{person}{Mari Ostendorf}, \bibinfo{person}{Luke Zettlemoyer}, \bibinfo{person}{Noah~A. Smith}, {and} \bibinfo{person}{Tao Yu}.} \bibinfo{year}{2023}\natexlab{}.
\newblock \showarticletitle{Binding Language Models in Symbolic Languages}. In \bibinfo{booktitle}{\emph{The Eleventh International Conference on Learning Representations}}.
\newblock
\urldef\tempurl%
\url{https://openreview.net/forum?id=lH1PV42cbF}
\showURL{%
\tempurl}


\bibitem[Console et~al\mbox{.}(2022)]%
        {fragmentsofbagrelationalalgebra}
\bibfield{author}{\bibinfo{person}{Marco Console}, \bibinfo{person}{Paolo Guagliardo}, {and} \bibinfo{person}{Leonid Libkin}.} \bibinfo{year}{2022}\natexlab{}.
\newblock \showarticletitle{Fragments of bag relational algebra: Expressiveness and certain answers}.
\newblock \bibinfo{journal}{\emph{Information Systems}}  \bibinfo{volume}{105} (\bibinfo{year}{2022}), \bibinfo{pages}{101604}.
\newblock
\showISSN{0306-4379}
\href{https://doi.org/10.1016/j.is.2020.101604}{doi:\nolinkurl{10.1016/j.is.2020.101604}}


\bibitem[Dai et~al\mbox{.}(2024)]%
        {uqe}
\bibfield{author}{\bibinfo{person}{Hanjun Dai}, \bibinfo{person}{Bethany~Yixin Wang}, \bibinfo{person}{Xingchen Wan}, \bibinfo{person}{Bo Dai}, \bibinfo{person}{Sherry Yang}, \bibinfo{person}{Azade Nova}, \bibinfo{person}{Pengcheng Yin}, \bibinfo{person}{Phitchaya~Mangpo Phothilimthana}, \bibinfo{person}{Charles Sutton}, {and} \bibinfo{person}{Dale Schuurmans}.} \bibinfo{year}{2024}\natexlab{}.
\newblock \showarticletitle{UQE: A Query Engine for Unstructured Databases}. In \bibinfo{booktitle}{\emph{Advances in Neural Information Processing Systems}}, \bibfield{editor}{\bibinfo{person}{A.~Globerson}, \bibinfo{person}{L.~Mackey}, \bibinfo{person}{D.~Belgrave}, \bibinfo{person}{A.~Fan}, \bibinfo{person}{U.~Paquet}, \bibinfo{person}{J.~Tomczak}, {and} \bibinfo{person}{C.~Zhang}} (Eds.), Vol.~\bibinfo{volume}{37}. \bibinfo{publisher}{Curran Associates, Inc.}, \bibinfo{pages}{29807--29838}.
\newblock
\urldef\tempurl%
\url{https://proceedings.neurips.cc/paper_files/paper/2024/file/34b3a40ec9752c1ae48fe85fef8fe8dc-Paper-Conference.pdf}
\showURL{%
\tempurl}


\bibitem[Delgado-Chaves et~al\mbox{.}(2025)]%
        {literaturescreening}
\bibfield{author}{\bibinfo{person}{Fernando~M. Delgado-Chaves}, \bibinfo{person}{Matthew~J. Jennings}, \bibinfo{person}{Antonio Atalaia}, \bibinfo{person}{Justus Wolff}, \bibinfo{person}{Rita Horvath}, \bibinfo{person}{Zeinab~M. Mamdouh}, \bibinfo{person}{Jan Baumbach}, {and} \bibinfo{person}{Linda Baumbach}.} \bibinfo{year}{2025}\natexlab{}.
\newblock \showarticletitle{Transforming literature screening: The emerging role of large language models in systematic reviews}.
\newblock \bibinfo{journal}{\emph{Proceedings of the National Academy of Sciences}} \bibinfo{volume}{122}, \bibinfo{number}{2} (\bibinfo{year}{2025}), \bibinfo{pages}{e2411962122}.
\newblock
\showeprint{https://www.pnas.org/doi/pdf/10.1073/pnas.2411962122}
\href{https://doi.org/10.1073/pnas.2411962122}{doi:\nolinkurl{10.1073/pnas.2411962122}}


\bibitem[Garcia-Molina et~al\mbox{.}(2000)]%
        {databasesystemimplementation}
\bibfield{author}{\bibinfo{person}{H. Garcia-Molina}, \bibinfo{person}{J.D. Ullman}, {and} \bibinfo{person}{J. Widom}.} \bibinfo{year}{2000}\natexlab{}.
\newblock \bibinfo{booktitle}{\emph{Database System Implementation}}.
\newblock \bibinfo{publisher}{Prentice Hall}.
\newblock
\showISBNx{9780130402646}
\showLCCN{99031049}


\bibitem[Grefen and de~By(1994)]%
        {amultisetextendedrelationalalgebra}
\bibfield{author}{\bibinfo{person}{P.W.P.J. Grefen} {and} \bibinfo{person}{R.A. de By}.} \bibinfo{year}{1994}\natexlab{}.
\newblock \showarticletitle{A multi-set extended relational algebra: a formal approach to a practical issue}. In \bibinfo{booktitle}{\emph{Proceedings of 1994 IEEE 10th International Conference on Data Engineering}}. \bibinfo{pages}{80--88}.
\newblock
\href{https://doi.org/10.1109/ICDE.1994.283002}{doi:\nolinkurl{10.1109/ICDE.1994.283002}}


\bibitem[Grumbach et~al\mbox{.}(1996)]%
        {querylanguagesforbags}
\bibfield{author}{\bibinfo{person}{St\'{e}phane Grumbach}, \bibinfo{person}{Leonid Libkin}, \bibinfo{person}{Tova Milo}, {and} \bibinfo{person}{Limsoon Wong}.} \bibinfo{year}{1996}\natexlab{}.
\newblock \showarticletitle{Query languages for bags: expressive power and complexity}.
\newblock \bibinfo{journal}{\emph{SIGACT News}} \bibinfo{volume}{27}, \bibinfo{number}{2} (\bibinfo{date}{July} \bibinfo{year}{1996}), \bibinfo{pages}{30–44}.
\newblock
\showISSN{0163-5700}
\href{https://doi.org/10.1145/235767.235770}{doi:\nolinkurl{10.1145/235767.235770}}


\bibitem[Grumbach and Milo(1993)]%
        {towardstractablealgebrasforbags}
\bibfield{author}{\bibinfo{person}{St\'{e}phane Grumbach} {and} \bibinfo{person}{Tova Milo}.} \bibinfo{year}{1993}\natexlab{}.
\newblock \showarticletitle{Towards tractable algebras for bags}. In \bibinfo{booktitle}{\emph{Proceedings of the Twelfth ACM SIGACT-SIGMOD-SIGART Symposium on Principles of Database Systems}} (Washington, D.C., USA) \emph{(\bibinfo{series}{PODS '93})}. \bibinfo{publisher}{Association for Computing Machinery}, \bibinfo{address}{New York, NY, USA}, \bibinfo{pages}{49–58}.
\newblock
\showISBNx{0897915933}
\href{https://doi.org/10.1145/153850.153855}{doi:\nolinkurl{10.1145/153850.153855}}


\bibitem[Guagliardo and Libkin(2017)]%
        {aformalsemanticsofsqlqueries}
\bibfield{author}{\bibinfo{person}{Paolo Guagliardo} {and} \bibinfo{person}{Leonid Libkin}.} \bibinfo{year}{2017}\natexlab{}.
\newblock \showarticletitle{A formal semantics of SQL queries, its validation, and applications}.
\newblock \bibinfo{journal}{\emph{Proc. VLDB Endow.}} \bibinfo{volume}{11}, \bibinfo{number}{1} (\bibinfo{date}{Sept.} \bibinfo{year}{2017}), \bibinfo{pages}{27–39}.
\newblock
\showISSN{2150-8097}
\href{https://doi.org/10.14778/3151113.3151116}{doi:\nolinkurl{10.14778/3151113.3151116}}


\bibitem[Liu et~al\mbox{.}(2025)]%
        {palimpzest}
\bibfield{author}{\bibinfo{person}{Chunwei Liu}, \bibinfo{person}{Matthew Russo}, \bibinfo{person}{Michael Cafarella}, \bibinfo{person}{Lei Cao}, \bibinfo{person}{Peter~Baile Chen}, \bibinfo{person}{Zui Chen}, \bibinfo{person}{Michael Franklin}, \bibinfo{person}{Tim Kraska}, \bibinfo{person}{Samuel Madden}, \bibinfo{person}{Rana Shahout}, {and} \bibinfo{person}{Gerardo Vitagliano}.} \bibinfo{year}{2025}\natexlab{}.
\newblock \showarticletitle{Palimpzest: Optimizing AI-Powered Analytics with Declarative Query Processing}. In \bibinfo{booktitle}{\emph{Proceedings of the {{Conference}} on {{Innovative Database Research}} ({{CIDR}})}} (2025).
\newblock


\bibitem[Liu et~al\mbox{.}(2024)]%
        {suql}
\bibfield{author}{\bibinfo{person}{Shicheng Liu}, \bibinfo{person}{Jialiang Xu}, \bibinfo{person}{Wesley Tjangnaka}, \bibinfo{person}{Sina Semnani}, \bibinfo{person}{Chen Yu}, {and} \bibinfo{person}{Monica Lam}.} \bibinfo{year}{2024}\natexlab{}.
\newblock \showarticletitle{{SUQL}: Conversational Search over Structured and Unstructured Data with Large Language Models}. In \bibinfo{booktitle}{\emph{Findings of the Association for Computational Linguistics: NAACL 2024}}, \bibfield{editor}{\bibinfo{person}{Kevin Duh}, \bibinfo{person}{Helena Gomez}, {and} \bibinfo{person}{Steven Bethard}} (Eds.). \bibinfo{publisher}{Association for Computational Linguistics}, \bibinfo{address}{Mexico City, Mexico}, \bibinfo{pages}{4535--4555}.
\newblock
\href{https://doi.org/10.18653/v1/2024.findings-naacl.283}{doi:\nolinkurl{10.18653/v1/2024.findings-naacl.283}}


\bibitem[Mittal et~al\mbox{.}(2024)]%
        {ssql}
\bibfield{author}{\bibinfo{person}{Akash Mittal}, \bibinfo{person}{Anshul Bheemreddy}, {and} \bibinfo{person}{Huili Tao}.} \bibinfo{year}{2024}\natexlab{}.
\newblock \bibinfo{title}{Semantic SQL -- Combining and optimizing semantic predicates in SQL}.
\newblock
\showeprint[arxiv]{2404.03880}~[cs.DB]
\urldef\tempurl%
\url{https://arxiv.org/abs/2404.03880}
\showURL{%
\tempurl}


\bibitem[Ong et~al\mbox{.}(2014)]%
        {DBLP:journals/corr/OngPV14}
\bibfield{author}{\bibinfo{person}{Kian~Win Ong}, \bibinfo{person}{Yannis Papakonstantinou}, {and} \bibinfo{person}{Romain Vernoux}.} \bibinfo{year}{2014}\natexlab{}.
\newblock \showarticletitle{The {SQL++} Semi-structured Data Model and Query Language: {A} Capabilities Survey of SQL-on-Hadoop, NoSQL and NewSQL Databases}.
\newblock \bibinfo{journal}{\emph{CoRR}}  \bibinfo{volume}{abs/1405.3631} (\bibinfo{year}{2014}).
\newblock
\showeprint[arXiv]{1405.3631}
\urldef\tempurl%
\url{http://arxiv.org/abs/1405.3631}
\showURL{%
\tempurl}


\bibitem[Patel et~al\mbox{.}(2025)]%
        {lotus}
\bibfield{author}{\bibinfo{person}{Liana Patel}, \bibinfo{person}{Siddharth Jha}, \bibinfo{person}{Melissa Pan}, \bibinfo{person}{Harshit Gupta}, \bibinfo{person}{Parth Asawa}, \bibinfo{person}{Carlos Guestrin}, {and} \bibinfo{person}{Matei Zaharia}.} \bibinfo{year}{2025}\natexlab{}.
\newblock \bibinfo{title}{Semantic Operators: A Declarative Model for Rich, AI-based Data Processing}.
\newblock
\showeprint[arxiv]{2407.11418}~[cs.DB]
\urldef\tempurl%
\url{https://arxiv.org/abs/2407.11418}
\showURL{%
\tempurl}


\bibitem[Shankar et~al\mbox{.}(2025)]%
        {docetl}
\bibfield{author}{\bibinfo{person}{Shreya Shankar}, \bibinfo{person}{Tristan Chambers}, \bibinfo{person}{Tarak Shah}, \bibinfo{person}{Aditya~G. Parameswaran}, {and} \bibinfo{person}{Eugene Wu}.} \bibinfo{year}{2025}\natexlab{}.
\newblock \bibinfo{title}{DocETL: Agentic Query Rewriting and Evaluation for Complex Document Processing}.
\newblock
\showeprint[arxiv]{2410.12189}~[cs.DB]
\urldef\tempurl%
\url{https://arxiv.org/abs/2410.12189}
\showURL{%
\tempurl}


\bibitem[Slivinskas et~al\mbox{.}(2001)]%
        {afoundationforconventionalandtemporalqueryoptimization}
\bibfield{author}{\bibinfo{person}{G. Slivinskas}, \bibinfo{person}{C.S. Jensen}, {and} \bibinfo{person}{R.T. Snodgrass}.} \bibinfo{year}{2001}\natexlab{}.
\newblock \showarticletitle{A foundation for conventional and temporal query optimization addressing duplicates and ordering}.
\newblock \bibinfo{journal}{\emph{IEEE Transactions on Knowledge and Data Engineering}} \bibinfo{volume}{13}, \bibinfo{number}{1} (\bibinfo{year}{2001}), \bibinfo{pages}{21--49}.
\newblock
\href{https://doi.org/10.1109/69.908979}{doi:\nolinkurl{10.1109/69.908979}}


\bibitem[Slivinskas et~al\mbox{.}(2002)]%
        {briningordertoqueryoptimization}
\bibfield{author}{\bibinfo{person}{Giedrius Slivinskas}, \bibinfo{person}{Christian~S. Jensen}, {and} \bibinfo{person}{Richard~Thomas Snodgrass}.} \bibinfo{year}{2002}\natexlab{}.
\newblock \showarticletitle{Bringing order to query optimization}.
\newblock \bibinfo{journal}{\emph{SIGMOD Rec.}} \bibinfo{volume}{31}, \bibinfo{number}{2} (\bibinfo{date}{June} \bibinfo{year}{2002}), \bibinfo{pages}{5–14}.
\newblock
\showISSN{0163-5808}
\href{https://doi.org/10.1145/565117.565119}{doi:\nolinkurl{10.1145/565117.565119}}


\bibitem[Zhuang et~al\mbox{.}(2025)]%
        {llmforpaperreview}
\bibfield{author}{\bibinfo{person}{Zhenzhen Zhuang}, \bibinfo{person}{Jiandong Chen}, \bibinfo{person}{Hongfeng Xu}, \bibinfo{person}{Yuwen Jiang}, {and} \bibinfo{person}{Jialiang Lin}.} \bibinfo{year}{2025}\natexlab{}.
\newblock \showarticletitle{Large language models for automated scholarly paper review: A survey}.
\newblock \bibinfo{journal}{\emph{Information Fusion}}  \bibinfo{volume}{124} (\bibinfo{year}{2025}), \bibinfo{pages}{103332}.
\newblock
\showISSN{1566-2535}
\href{https://doi.org/10.1016/j.inffus.2025.103332}{doi:\nolinkurl{10.1016/j.inffus.2025.103332}}


\end{thebibliography}
